%% file: apssamp.tex
\documentclass[%
 reprint,
superscriptaddress,
nofootinbib,
 amsmath,amssymb,
 aps,
]{revtex4-2}

\usepackage{graphicx}
\usepackage{dcolumn}
\usepackage{bm}

\usepackage[dvipsnames]{xcolor}
\usepackage{float}

\usepackage{lineno}

\usepackage{hyperref}

\begin{document}

\preprint{APS/123-QED}

\title{The DAMA/LIBRA signal: an induced modulation effect?}

\input{author_list}

\begin{abstract}
The persistence of the DAMA/LIBRA (DAMA) modulation over the past two decades has been a source of great contention within the dark matter community. The DAMA collaboration reports a persistent, modulating event rate within their setup of NaI(Tl) scintillating crystals at the INFN Laboratori Nazionali del Gran Sasso (LNGS) underground laboratory \cite{Bernabei_2021}. A recent work alluded that this signal could have arisen due to an analysis artefact, caused by DAMA not accounting for time variation of decaying background radioisotopes in their analysis procedure \cite{Adhikari_2023}. In this work, we examine in detail this `induced modulation' effect, arguing that a number of aspects of the DAMA signal are incompatible with an induced modulation arising from decays of background isotopes over the lifetime of the experiment. Using a toy model of the DAMA/LIBRA experiment, we explore the induced modulation effect under different variations of the activities of
the relevant isotopes -- namely, $^3$H and $^{210}$Pb -- highlighting the various inconsistencies between the resultant toy datasets and the DAMA signal. We stress the importance of the SABRE experiment, whose goal is to unambiguously test for the presence of such a modulating signal in an experiment using the same target material and comparable levels of background~\cite{Antonello_2019}.
\end{abstract}

\maketitle

The DAMA/LIBRA modulation, persisting with an amplitude of 0.00959 $\pm$ 0.00076 cpd/kg/keV$_{\text{ee}}$ at 12.6$\sigma$ significance over approximately 16 years in the 2-6~keV$_{\text{ee}}$\footnote{Henceforth, energies quoted in keV can be assumed to be keV$_{\text{ee}}$ unless explicitly stated otherwise.} energy region \cite{Bernabei_2021}, remains unexplained in the eyes of most within the dark matter community. Whilst attempts have been made to reconcile this signature with spin-independent or spin-dependent WIMP dark matter scattering \cite{Savage_2009}, other experimental efforts have excluded the parameter space that would produce such a signal by many orders of magnitude. This includes the strongest current constraints from experiments that use multi-tonne liquid xenon targets~\cite{Aalbers_2023, Aprile_2023, Meng_2021}, in addition to constraints from experiments using liquid argon~\cite{Ajaj_2019} and solid state~\cite{Hehn_2016} targets. Even when the full space of non-relativistic effective field theory WIMP operators is considered \cite{Kang_2018}, tension between the best-fit solutions and current experimental constraints \cite{Aalbers_2024} is found.

Ongoing efforts are underway from a number of experimental collaborations aiming to unambiguously confirm or refute the DAMA signal in experiments using the same target material, including already published initial results from the COSINE-100~\cite{Adhikari_2022} and ANAIS-112~\cite{Amare_2021} experiments. Such a determination is of great importance; to date, no satisfactory explanation for the signal consistent with other experimental results has been given. Recent works have debated whether aspects of the modulation reported by DAMA could be explained by a particular analysis technique that they apply to their data~\cite{Adhikari_2023, Messina_2020}. Here, we show that a more detailed exploration of this effect reveals strong inconsistencies between the signal-like component `induced' by the analysis technique and the results reported by DAMA. In particular, we highlight the incorrect phase of the induced signal, the substantial reduction in amplitude coming from a re-assessment of their likely backgrounds, and the non-constant nature of the signal amplitude in time. We additionally explore an alternative statistical treatment of the data used by DAMA, showing that whilst this also has the potential to produce a bias under their assumptions on their background model, the bias produced is once again inconsistent with the reported results. Finally, we examine the interplay between the DAMA analysis technique and a genuinely modulating component, be it from a dark matter signal or a source of background.

In a similar vein to a previous study of this induced modulation effect~\cite{Adhikari_2023}, and in the absence of a detailed background model from DAMA, we use the SABRE South NaI(Tl) crystal background model~\cite{Barberio_2023} to build a toy background model for the DAMA experiment. The SABRE South experiment is currently undergoing commissioning at the Stawell Underground Physics Laboratory in Victoria, Australia. It will consist of 50 kg of ultra-pure NaI(Tl) crystals in 7 modules, each coupled to two PMTs. The experiment will be operated at room temperature, with the crystal modules surrounded by 12 kl of linear alkylbenzene (LAB) liquid scintillator to act as an active veto system. On top of the steel shielding surrounding the veto vessel will be 9.6 m$^2$ of EJ200 scintillators for active rejection of muon-induced events. It is part of a dual-site NaI(Tl) experiment whose primary goal is to provide an independent test of the DAMA/LIBRA signal~\cite{Antonello_2019}. Its dual-site nature, with one detector in the Southern Hemisphere and one in the Northern Hemisphere at LNGS, will allow for strong discrimination against seasonal systematic effects. By using ultra-pure crystals with a lower background than those used in DAMA/LIBRA, SABRE aims to provide a conclusive confirmation or refutation of the DAMA/LIBRA in 2-3 years of runtime \cite{Barberio_2023}.

We find that the induced modulation effect comes from the application of the DAMA analysis method in the presence of a background rate that is not constant in time. As identified in Ref.~\cite{Adhikari_2023}, the reported rate of single-hit events in the DAMA experiment was found to decrease between phase 1 and phase 2 of the experiment\footnote{Phase 1 ran from September 2003 to August 2010, whilst phase 2 ran from November 2011 to October 2019, with PMT upgrades occuring in between.}~\cite{Bernabei_2008, Bernabei_2018}. Background modelling from the NaI(Tl) experiments COSINE-100~\cite{Adhikari_2021} and ANAIS-112~\cite{Amare_2019}, as well as background simulation from the SABRE South experiment~\cite{Barberio_2023}, suggests that time variation in the 2-6~keV DAMA region of interest (ROI) over the course of the experiment\footnote{Whilst other cosmogenically activated isotopes do exhibit time variation, these timescales are either much longer or much shorter than the experimental runtime.} is expected from the decay of the radioisotopes $^3$H (tritium, half-life 12.3 years) and $^{210}$Pb (half-life 22.3 years). As such, we scale the initial activities of these background model components from the values used for the SABRE South simulation to match the upper limits reported on each by DAMA~\cite{Bernabei_2008_2}. These are 0.09 and 0.03 mBq/kg for $^3$H and $^{210}$Pb, respectively. In addition, we include a component of $^{210}$Pb within the Teflon wrap of the crystals. This is a conservative choice given some evidence from the characterisation of the SABRE North crystal, denoted NaI-33, that this activity could be higher than previously assumed \cite{Mariani_2022, Barberio_2023}, although we note that its inclusion has little impact. The upper limit on the $^{210}$Pb activity from the SABRE North measurement is used to scale this particular surface component in the SABRE South simulation, correcting for the different crystal surface area-to-volume ratios. All other initial activities are set to the values reported in the SABRE South simulation paper~\cite{Barberio_2023}. Given that the induced modulation effect comes from the decaying components, we consider this assumption to be reasonable. Our assumption is further supported by the expected similarity in crystal purity and composition between the DAMA crystals and the SABRE South crystals~\cite{Bernabei_2020, Calaprice_2022}. We do not consider the effect of muon-induced neutron backgrounds in this work~\cite{Klinger_2015}. .

Using our DAMA background model, toy DAMA datasets can be generated via a Monte Carlo (MC) procedure. Each component has its expected number of events calculated based on the full DAMA exposure summed across phase 1 and phase 2 of the experiment, details of which are found in Refs.~\cite{Bernabei_2013, Bernabei_2021}. In calculating this from the initial activities, the decay of each component over the runtime of the experiment is accounted for. Each component then has single-hit crystal energy depositions drawn from the single-hit energy spectra of the SABRE South background model, with the number of depositions drawn allowed to fluctuate according to Poisson statistics. As DAMA/LIBRA does not contain analogous veto systems to those in SABRE South, we do not include veto information in this toy model. The drawn energies are then transformed stochastically via a Gaussian smearing process to reflect the finite energy resolution of the DAMA detector. This smearing is applied with the measured DAMA resolution $\sigma_E = 0.488 \sqrt{E} + 0.0091 E$, where $E$ is the energy of the true deposition in units of keV \cite{Bernabei_2008_2}. The energy-dependent efficiency of the detector, which encodes information such as the trigger requirement that a photoelectron is observed in both crystal PMTs in coincidence, has been extracted from Ref.~\cite{Bernabei_2012} for both phases of the experiment.

Each smeared energy drawn in a given iteration of the MC undergoes a Bernoulli trial with success probability equal to the detector efficiency at that energy, determining whether the event is kept in that MC dataset. Of the remaining events, we keep only those with smeared energy lying within in 2-6~keV ROI. We do not attempt to model the effect of non-linearity in the scintillation response, instead using the smeared energy directly as a proxy for a linearity-corrected reconstructed energy. Given that such non-linearities are observed to be small in the DAMA crystals, this choice is justified \cite{Bernabei_2018}. Each surviving MC event from each background component then has an event time randomly drawn according to the exposure distribution throughout each so-called annual cycle\footnote{The cycles are all close to 1 year in length, but with substantial variations of up to 3 months.} in phases 1 and 2 of the experiment, adjusted for the decaying activity of that particular component throughout the runtime. This output is combined across all background components to create a single realisation of a toy dataset.

\begin{figure}[H]
    \includegraphics[scale=0.55]{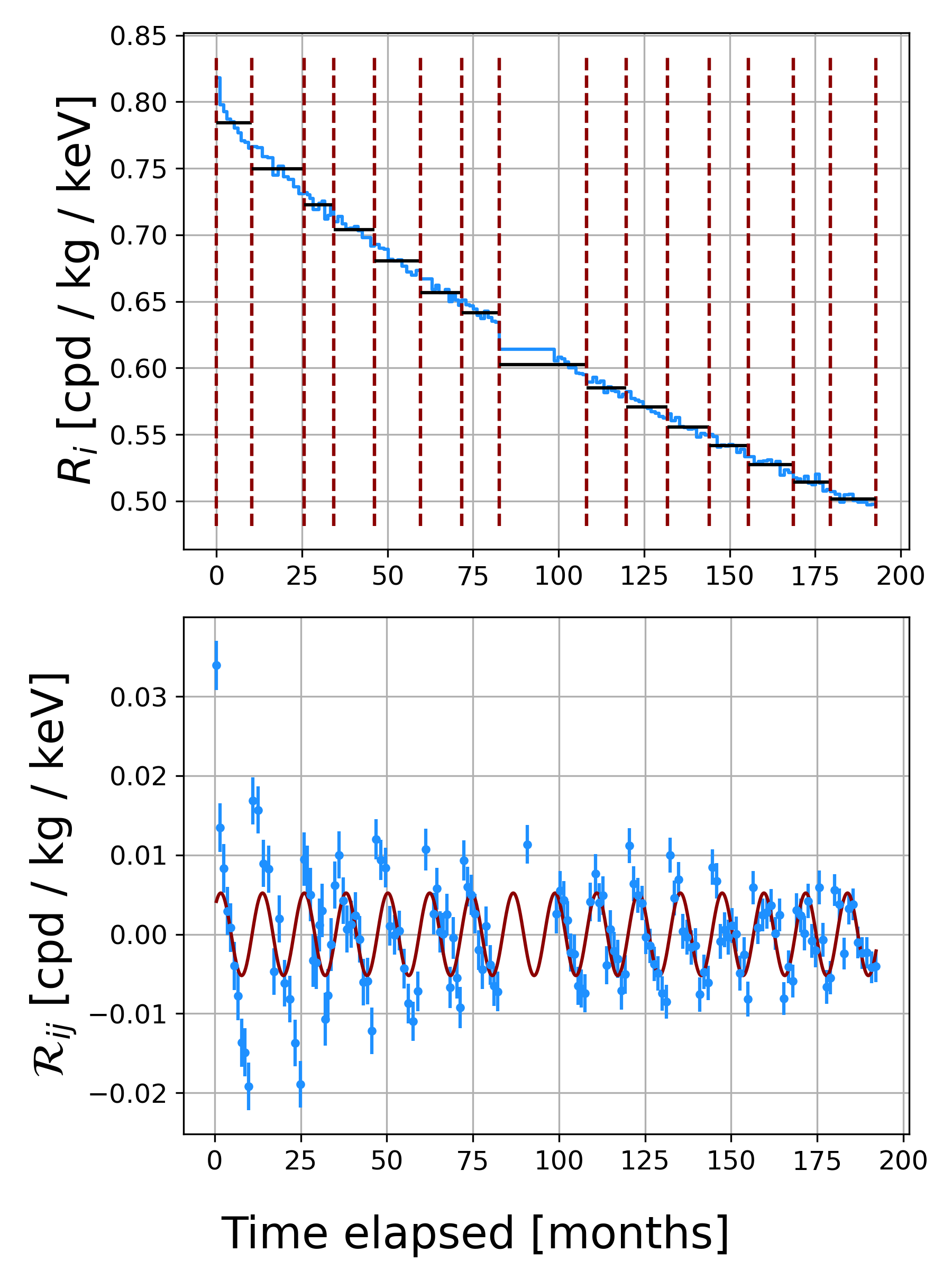}
    \caption{\label{fig:toy_dataset_residuals}\textbf{Top:} a single realisation of a toy DAMA/LIBRA dataset, generated as described in the text over the entire experimental exposure. The exposure-normalised number of counts in 10 bins across each approximately annual cycle -- depicted with dashed vertical lines -- is plotted in red. Black horizontal lines denote the average of the normalised rate over each annual cycle. Time bins/annual cycles are plotted such that they appear continuous/contiguous (see text). \textbf{Bottom:} residuals distribution for the above dataset, calculated for each bin as described in the text. The red line shows the result of fitting an oscillating function to the residuals, as described in the text. Error bars on the residuals are calculated by combining the Poisson errors on the counts in each bin with the error on the mean counts subtracted within each annual cycle.}
\end{figure}

We show one such realisation of these toy datasets in Fig.~\ref{fig:toy_dataset_residuals}. We follow the methodology of DAMA~\cite{Bernabei_2021}, dividing the data into 10 time bins within each approximately annual cycle. The precise details of these cycles are described in Refs.~\cite{Bernabei_2013, Bernabei_2021}; there are seven in phase 1 and eight in phase 2, with a gap of approximately 1 year and 3 months between them when the detector was upgraded\footnote{The crystals remained identical, with the main change being a PMT upgrade \cite{Bernabei_2012}. As the background contribution from PMT radioactivity is so subdominant to that from crystal radiogenic and cosmogenic isotopes 
\cite{Barberio_2023}, we do not consider this here.} \cite{Bernabei_2012}. In plotting these, we concatenate the second through to last bin edges within each annual cycle to the full set of bin edges from the first annual cycle to depict a continuous series of time bins. We also concatenate the endpoint of each annual cycle to the start point and endpoint of the first annual cycle to depict the annual cycle divisions as contiguous. For each annual cycle, the normalised rate, denoted $R_i$ for the $i$th time bin, is plotted. It is defined as
\begin{equation}
    R_i = \frac{N_i}{\mathcal{E}_i (E_{\text{max}} - E_{\text{min}})},
\end{equation}
where $N_i$ is the number of counts in that bin, $\mathcal{E}_i$ is the exposure (in kg days) of that particular time bin, and $E_{\text{min}}$ and $E_{\text{max}}$ define the lower and upper endpoints of the ROI, respectively. In the process of obtaining $N_i$, events are first binned in a two dimensional space $N_{ij}$, where $j$ denotes energy bins of width 0.5~keV. The counts in each bin are then scaled to correct for the detection efficiency at the centre of that bin, and $N_{ij}$ is projected onto the time axis to obtain $N_i$. This follows the procedure taken by DAMA~\cite{Bernabei_2008}.

Within each annual cycle, denoted $j$, the DAMA methodology is to calculate the so-called residual rate, $\mathcal{R}_{ij}$. This is defined as
\begin{equation}
    \mathcal{R}_{ij} = R_i - \langle R_i \rangle_j,
\end{equation}
where $\langle R_i \rangle_j$ denotes the event rate averaged over all bins of the $j$th cycle. We plot these residuals for the same toy dataset in Fig.~\ref{fig:toy_dataset_residuals}. The DAMA methodology\footnote{DAMA also performs fits with a fixed period and phase; throughout this work, we employ a fit where all three parameters freely float.} then fits a modulating function of the form $\mathcal{A} \cos\left[\frac{2\pi}{T} t_i + \phi\right]$ to the ${R}_{ij}$, with $t_i$ being the time at the centre of the $i$th time bin and $\mathcal{A}$, $\phi$ and $T$ being the amplitude, phase and period of the modulation, respectively. The motivation behind this approach is to provide a measure of the modulating event rate around a constant background within each annual cycle. However, as is evident from Fig.~\ref{fig:toy_dataset_residuals}, in the presence of a time-varying background this method produces a bias. This is the induced modulation effect; where the background rate is monotonically changing with time, such a procedure will produce a sawtooth-like set of residuals. The period  of these residuals is determined by the length of the averaging period, the amplitude by the magnitude of the variation of the rate around the average within a cycle, and the phase by the nature of the time variation with respect to the average. We show the result of performing this fit to this particular set of residuals in Fig.~\ref{fig:toy_dataset_residuals}.

\begin{figure*}
    \includegraphics[scale=0.6]{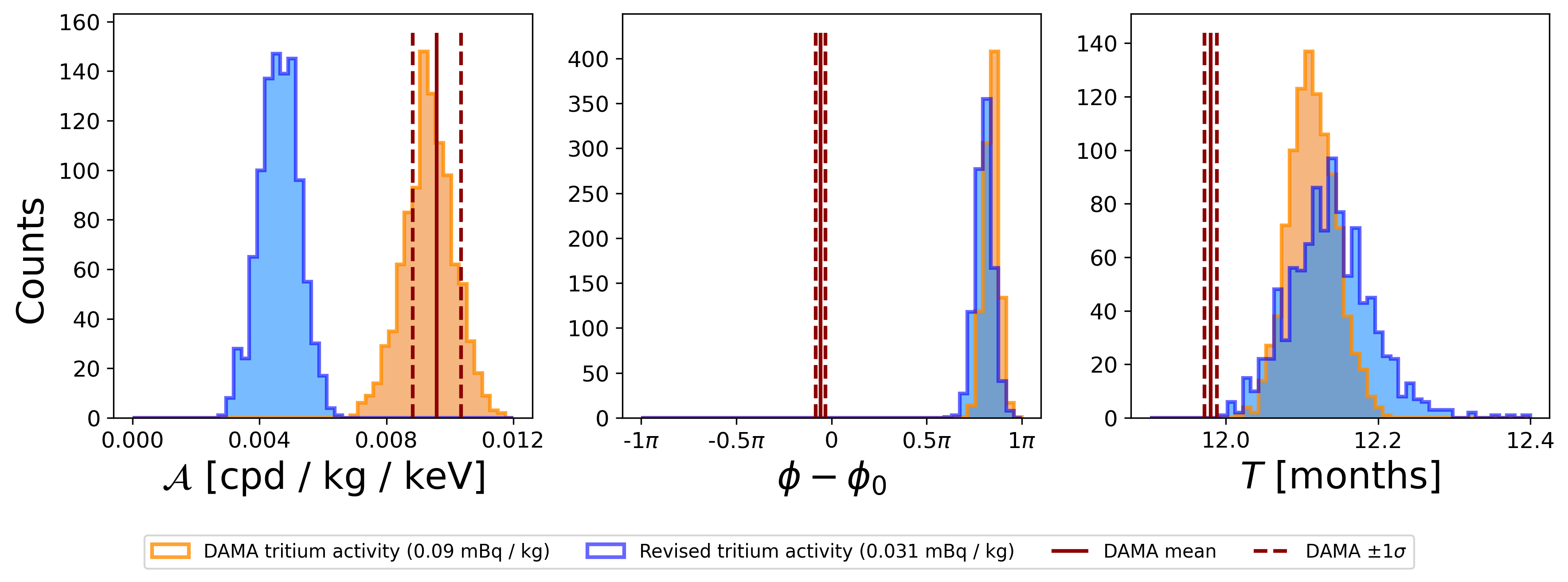}
    \caption{\label{fig:all_toys_residuals}Results of 1000 toy residuals fits to MC realisations of the DAMA/LIBRA dataset, as described in the text. We show the distributions of amplitude, phase and period of the oscillating fit function for two different assumptions on the initial activity of $^3$H. These are the `DAMA tritium activity' corresponding to the DAMA-reported upper limit of 0.09 mBq/kg, and the `revised tritium activity' corresponding to the 0.031 mBq/kg initial activity estimate discussed in the text. For comparison we show the results reported by DAMA in each case \cite{Bernabei_2021}.}
\end{figure*}

Following a similar approach to Ref.~\cite{Adhikari_2023}, we study the expected distribution of amplitudes, phases, and periods under the application of this analysis procedure over a large number (1000) of toy realisations of the DAMA dataset. Each dataset is generated over the 15 annual cycles across DAMA/LIBRA phases 1 and 2, with the ROI being restricted to 2-6~keV for comparison with the combined residuals analysis presented by DAMA \cite{Bernabei_2021}. Above the energy threshold of 2~keV\footnote{1~keV for DAMA/LIBRA phase 2.}, DAMA state that they have full rejection of background events from PMT noise \cite{Bernabei_2020}, and as such we do not consider this component here. For each toy, a fit of the form $\mathcal{A} \cos\left[\frac{2\pi}{T} t_i + \phi\right]$ is performed as a $\chi^2$ minimisation using the residuals ${R}_{ij}$ and their uncertainties. We show the resultant distributions of the three fit parameters in orange in Fig.~\ref{fig:all_toys_residuals}. Similarly to Ref.~\cite{Adhikari_2023}, we find that the amplitudes $\mathcal{A}$ determined from the fits are clustered around the amplitude reported by DAMA, whilst the periods $T$ are clustered around 12.1 months, just above the fit value of 12 months reported by DAMA~\cite{Bernabei_2021}. However, crucially, we find that the phases $\phi$ determined from the fits -- relative to the expected dark matter modulation phase $\phi_0$ corresponding to June 2nd -- are clustered close to $\pi$, i.e. they are close to being in antiphase with the modulation signal that would originate from a galactic dark matter source. A similar effect was observed in Ref.~\cite{Adhikari_2023}, and we emphasise here its importance: such an induced modulation effect cannot explain the result of the fit presented by DAMA. Given the fixed locations of the annual cycles used for the averaging procedure, an induced modulation effect giving an approximately consistent phase with the DAMA fit could only come from a monotonically increasing background rate, which seems unphyiscal.

Whilst this conclusion alone provides strong evidence that this effect cannot explain the DAMA result, it warrants further scrutiny. In particular, the upper limit reported by DAMA on the activity of the radioisotope $^3$H -- produced in the NaI(Tl) crystals by cosmogenic activation \cite{Villar_2018} -- appears to be well above the value we infer from our own attempt to estimate the initial $^3$H activity based on publicly available information about the DAMA/LIBRA experiment. In deriving this estimation, we make a number of assumptions that lead to our result being an overly conservative upper limit estimation of the activity, but we still find a lower activity than that reported by DAMA. To obtain this estimate, we take the result from a recent detailed calculation that the $^3$H production rate in an NaI(Tl) crystal is 83 $\pm$ 27 atoms/kg/day, normalised to the reference location of New York City at sea level~\cite{Amare_2018}. Here, we take the +1$\sigma$ value of 110 atoms/kg/day. It is known that the DAMA/LIBRA crystals were grown by the Kyropoulos method~\cite{Bernabei_2008_2}. It has been demonstrated that crystals as large as 120 kg can be grown by this method in 3-4 days \cite{Park_2020}, so we conservatively attribute 5 days to the production of each of the 25 $\times$ 10 kg crystals used by DAMA. Under the (again conservative) assumption that the crystals are grown sequentially and are not shipped until all production is completed, this gives 125 days of surface exposure during crystal growth. In reality, it seems likely that DAMA would have made multiple shipments, as they report taking steps to ensure that the crystals spent as little time above ground as possible \cite{Bernabei_2008_2}. The DAMA crystals were produced by Saint-Gobain \cite{Sutanto_2023}, now Luxium Solutions. They have two production facilities in Europe; one in Nemours (France) and another in Gières (France) \cite{Luxium}, by far the closest facilities to LNGS. As such, it seems reasonable to assume that the crystals were produced at one of these two facilities. DAMA reports that the crystals were transported by road to LNGS \cite{Bernabei_2008_2}; therefore, we make another conservative attribution of 5 days of further surface exposure during transport.

Before calculating our estimation of the initial $^3$H activity in the DAMA crystals using this estimate of surface exposure, we consider variation in the cosmogenic activation rate due to variations in altitude and geomagnetic shielding. Following calculations in Ref.~\cite{Pettus_2015}, it can be shown that a reference rate of cosmogenic activation calculated for New York City at sea level can be modified to another location using the factor $F_{\text{alt}} F_{\text{GS}}$. The factor $F_{\text{alt}}$ reflects the variation due to air pressure, and the factor $F_{\text{GS}}$ the variation due to geomagnetic shielding. They can be calculated as~\cite{Pettus_2015}
\begin{subequations}
\begin{align}
    F_{\text{alt}} &= \exp\left[ \frac{d_{\text{ref}} - d}{L_n} \right]  \\
    F_{\text{GS}} &= N \left[ 1 - \text{exp} \left( \frac{-\alpha(P)}{R_c^{k(P)}} \right) \right].
\end{align}
\end{subequations}
Here, $d$ is a scaling of the atmospheric pressure $P$, with $d_{\text{ref}}$ corresponding to the pressure at sea level, while $L_n$ is a constant related to the attenuation length of neutrons above 10~MeV, weighted by atmospheric density. The functions of pressure $\alpha(P)$ and $k(P)$ are detailed in Ref.~\cite{Pettus_2015}. The normalisation constant $N$ accounts for the reference location of New York City at sea level. Finally, $R_c$ denotes the vertical geomagnetic cutoff rigidity, a measure of the geomagnetic shielding effect, which can be calculated using various publicly available tools. Here, we used the tool found at Ref.~\cite{Cutoff}. We calculate a correction due to the factor $F_{\text{alt}}$, which accounts for both atmospheric pressure at the production site and the variation in pressure during transport to LNGS along the likely transportation routes \cite{route_plotter}. Both production locations of Nemours and Gières give a similar estimation of the magnitude of this correction. Calculations of $R_c$ for each location yield estimations of the factor $F_{\text{GS}} \approx 0.98$ for Nemours and $F_{\text{GS}} \approx 0.93$ for Gières.

Conservatively then, we factor in the increase in $^3$H production rate from the factor $F_{\text{alt}}$ and neglect the reduction due to the geomagnetic shielding factor $F_{\text{GS}}$. This allows us to calculate our estimate of the initial DAMA $^3$H activity as 0.031 mBq/kg, substantially lower than the 0.09 mBq/kg reported as an upper limit by DAMA. We additionally neglect any so-called `cooldown' period, where the reduction in $^3$H activity between the crystals being brought underground and the start of DAMA/LIBRA phase 1 would result in a lower initial activity than that calculated here. This ensures that we reach an upper limit on the likely activity of tritium at the start of the experiment.

We now study the impact of this lower $^3$H background on the induced modulation effect, by performing the same series of toy fits as before, this time setting the initial $^3$H activity in the toy background model to 0.031 mBq/kg. We show the resultant distributions of $\mathcal{A}$, $\phi$ and $T$ in Fig.~\ref{fig:all_toys_residuals}, this time in blue. Whilst the distributions of both the period and the phase determined in the fits are similar to before, we find the distribution of amplitudes to be shifted to much lower values, by a factor $\sim$2. This reflects the relative rate reduction across each annual cycle being lower as a consequence of the initial $^3$H activity being lower. We view this as further evidence against the hypothesis that the DAMA signal is an artefact of their determination of the modulation; with what we see as a more likely, but still over-attributed estimation of their $^3$H background, neither the amplitude nor the phase of the induced modulation are consistent with their reported results. We note that both $^3$H activity assumptions lead to a poor goodness of fit metric: the mean $\chi^2 / d.o.f.$ values over the 1000 toys are found to be 471 / 147 and 413 / 147 for the DAMA-reported and revised activities, respectively. This can be compared with 130 / 155 for DAMA's fit \cite{Bernabei_2021}\footnote{It should be noted that this is for a fit with a fixed period and phase, and also includes data from DAMA/NaI.}. The fact that a better fit is found for the revised $^3$H activity case can be attributed to the fact that there is a less dramatic decrease in modulation amplitude in this case.

Next, we examine an alternative methodology employed by DAMA to present their results. This involves performing a maximum likelihood fit for different energy bins. For an observed number of counts $N_{ij}$ in the $i$th time bin and the $j$th energy bin, the Poissonian likelihood $L_j$ for the $j$th energy bin is
\begin{equation}
    L_j = \prod_i \frac{e^{-\mu_{ij}}\mu_{ij}^{N_{ij}}}{N_{ij}!},
\end{equation}
where $\mu_{ij}$ is the expected number of events in that particular bin. Their fitting methodology is to model this as
\begin{equation}\label{eqn:mu_ij}
    \mu_{ij} = \left[ c_j + m_j \cos\left(\frac{2\pi}{T} (t_i - t_0)\right) \right] \mathcal{E}_i \Delta E_j \epsilon_j.
\end{equation}
Here, $\mathcal{E}_i$ is the exposure of the $i$th time bin, $\Delta E_j$ is the width of the $j$th energy bin and $\epsilon_j$ is the energy-dependent efficiency at the centre of that energy bin. The time at the centre of the $i$th time bin is given by $t_i$, $T$ is the period of the oscillating component -- fixed here to 365.25 days -- and $t_0$ is the time of maximum modulation amplitude, fixed here to 152.5 days. The free parameters of the fit are $c_j$ and $m_j$, controlling the amplitude of a constant and modulating component for each energy bin.

A maximum likelihood fit is performed for each $L_j$, giving the maximum likelihood estimators $\hat{c}_j$ and $\hat{m}_j$. DAMA/LIBRA reported the values of $\hat{m}_j$ in order to quantify the variation in the amplitude of the modulation as a function of energy~\cite{Bernabei_2021}. DAMA argue that $\hat{m}_j$ being greatest within the lowest-energy bins and much lower in the higher-energy bins supports the hypothesis that the modulation is due to a WIMP-like signal, given that such a modulation has greater concentration in the lower-energy bins. In order to examine the effect of time variation in background counts on this fitting methodology, we once again generate 1000 toy datasets and perform these maximum likelihood fits for each. The datasets are generated assuming the DAMA-reported upper limit on $^3$H activity, as well as our alternative estimate outlined earlier. For consistency with DAMA, we choose 1-day time bins and energy bins of width 0.5~keV \cite{Bernabei_2021}.

\begin{figure}[H]
    \includegraphics[scale=0.52]{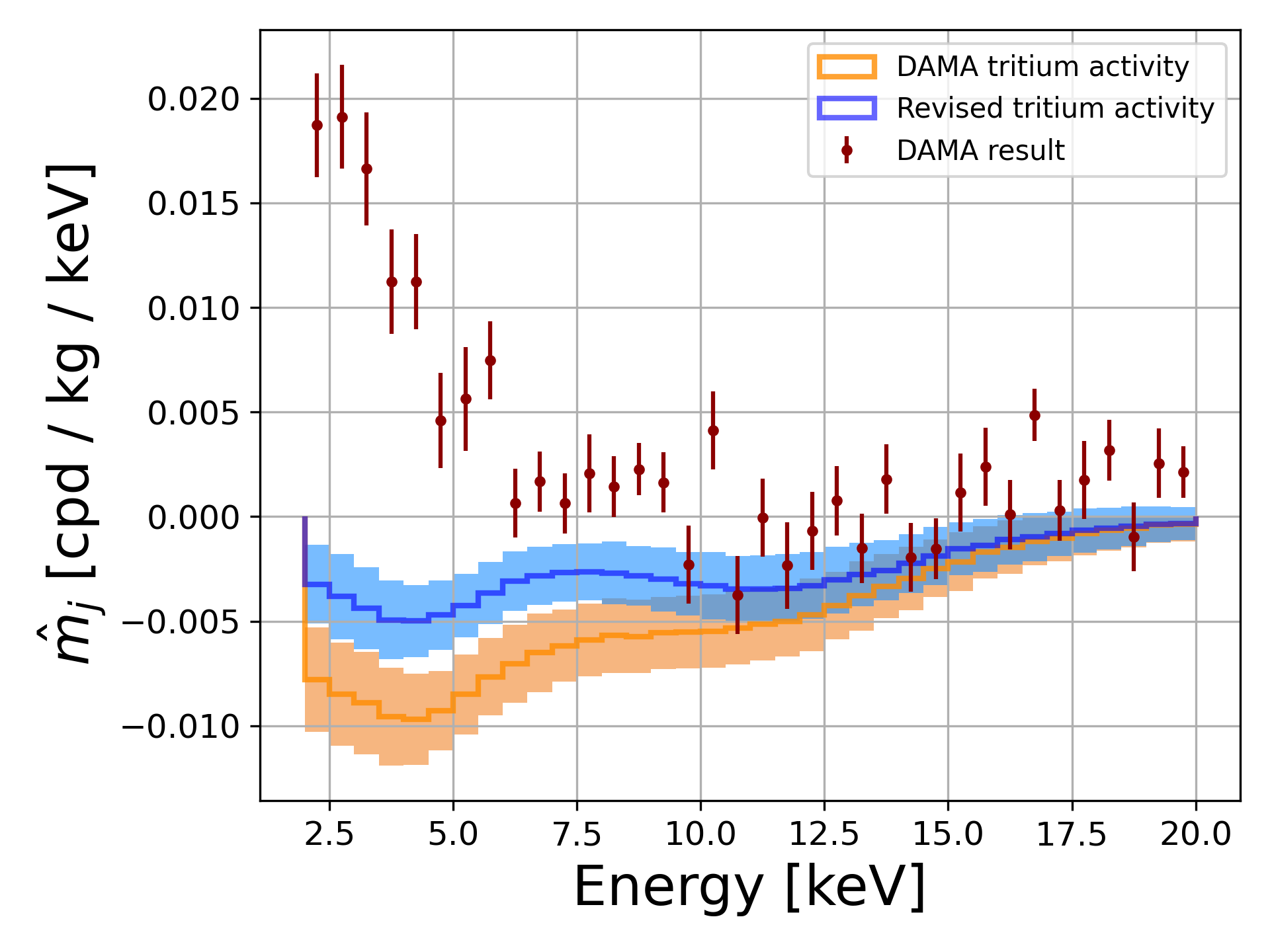}
    \caption{\label{fig:all_toys_energy_bins}Results of 1000 MC realisations of the maximum likelihood fit described in the text, performed over energy bins. For each energy bin, the mean and $\pm1\sigma$ values of the fitted parameter $\hat{m}_j$ are shown. The same two scenarios for the initial activity of $^3$H as in Fig. \ref{fig:all_toys_residuals} are explored. We compare the results of these toy fits to the result of the same fit as reported by DAMA \cite{Bernabei_2021}.}
\end{figure}

In Fig.~\ref{fig:all_toys_energy_bins} we show the values of $\hat{m}_j$ determined in the toy fits for both $^3$H activity assumptions, as well as the DAMA-reported results, in the 2-20~keV energy range. Whilst a decay in amplitude of $\hat{m}_j$ at higher energies is seen in all cases, the initial amplitude is much smaller in our MC-derived results versus the DAMA result, this initial amplitude being smaller for the lower $^3$H activity case. Crucially, the sign of $\hat{m}_j$ in these lower energy bins is opposite to that reported in the DAMA fits. The inconsistency between what we see here and the reported DAMA results -- even for our alternative estimation of $^3$H activity -- suggests that the DAMA signal cannot be attributed to an induced modulation effect coming from the decay of $^3$H and $^{210}$Pb over the runtime of the experiment.

\begin{figure}[H]
    \includegraphics[scale=0.52]{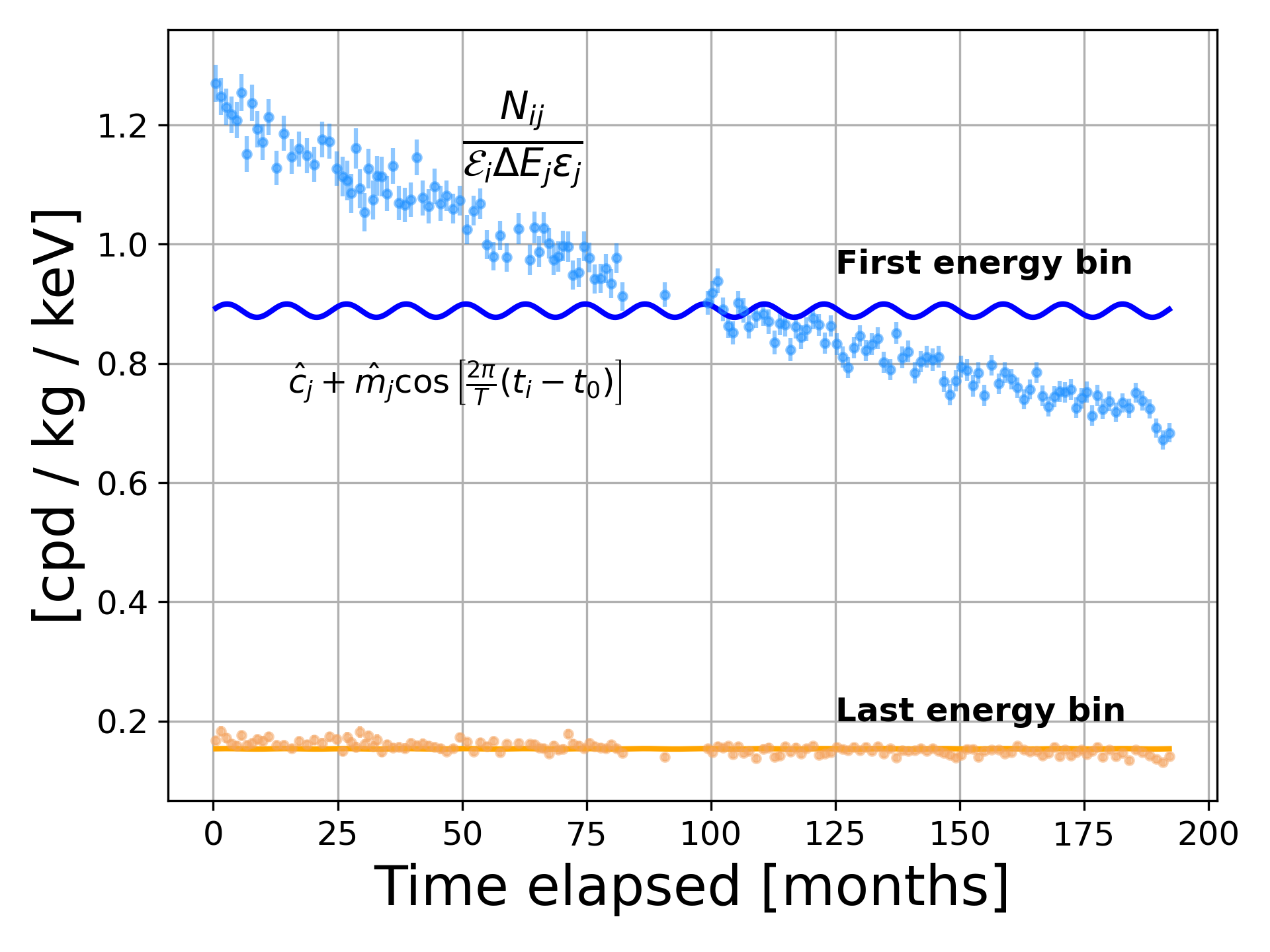}
    \caption{\label{fig:toy_dataset_energy_bins}A single realisation of a toy DAMA/LIBRA dataset, as in Fig. \ref{fig:toy_dataset_residuals}. This time we show the distribution in counts $N_{ij}$ across the $i$ time bins and for two choices of energy bins $j$ in the 2-20~keV ROI: the first (2-2.5~keV) and the last (19.5-20~keV). The counts are normalised by exposure, energy bin width and efficiency, and compared with the function used to calculate the expected number of counts in each bin, evaluated using the maximum likelihood estimators of its parameters.}
\end{figure}

It is interesting to observe where the energy dependence of this fit in the presence of such high concentrations of these isotopes comes from. Given that there is no notion of yearly-averaged background subtraction present in this fitting methodology, it would be misleading to attribute what we are seeing to the same induced modulation effect seen with the residuals. To understand why an energy dependence is observed, it is helpful to visualise the distribution of the obeserved counts $N_{ij}$ and the expected counts $\mu_{ij}$. Figure \ref{fig:toy_dataset_energy_bins} shows normalised $N_{ij}$ and $\mu_{ij}$, evaluated at the best-fit values $\hat{c}_j$ and $\hat{m}_j$, for both the first and last energy bins, for a particular realisation of a toy dataset using the DAMA-reported upper limit on $^3$H activity. As can be seen, a strong decay in counts across time bins is seen for the first energy bin, with an almost imperceptible level of decay for the last energy bin. We attribute this to the fact that the energy spectra of the components with the strongest reduction in count rates across the live time of the experiment -- $^3$H and $^{210}$Pb -- are concentrated in the lowest-energy bins. In neither case is a modulation seen. We see that $\hat{m}_j$ taking a non-zero and negative value in the first energy bin is simply a consequence of this value minimising the deviation between observed and expected counts across bins.

\begin{figure}[H]
    \includegraphics[scale=0.52]{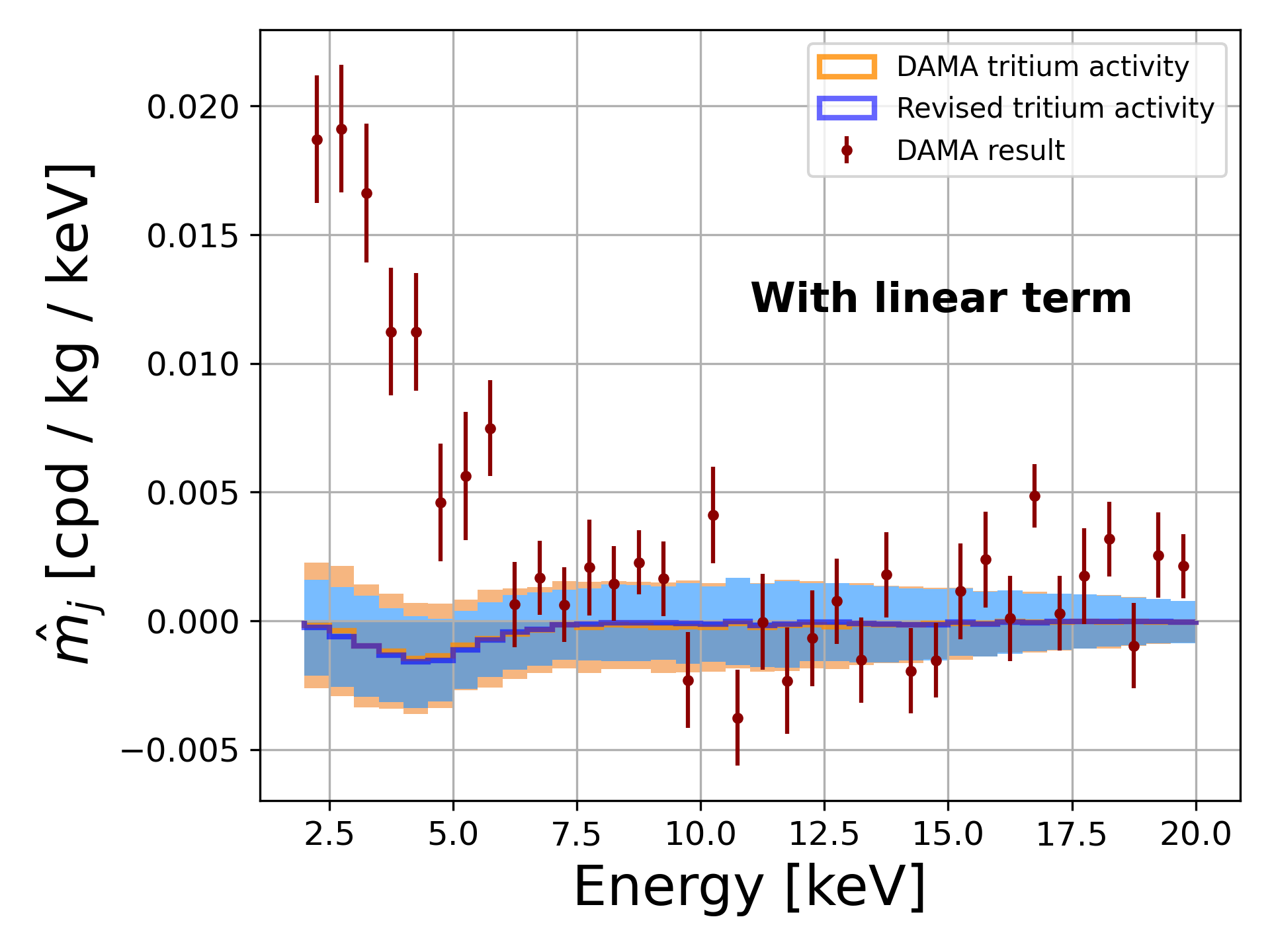}
    \caption{\label{fig:all_toys_energy_bins_linear_term}The result of performing the same analysis as in Fig. \ref{fig:all_toys_energy_bins}, this time including a linear term in Eqn. \ref{eqn:mu_ij}. We once again compare with the DAMA-reported results \cite{Bernabei_2021}, unchanged with the inclusion of the linear term \cite{Bernabei_2020}.}
\end{figure}

We note that DAMA considered including a non-oscillatory time dependence in this approach by including in the square brackets of Eqn. \ref{eqn:mu_ij} a term $l_j t_i$, i.e. a linear decrease (or increase) in rate over time in each energy bin, finding little change in their fit results \cite{Bernabei_2020}. We repeat the analysis procedure used to produce Fig. \ref{fig:all_toys_energy_bins}, this time including this linear component in our fit function for $\mu_{ij}$. We show the result of this in Fig. \ref{fig:all_toys_energy_bins_linear_term}, where it is clear that we still see a strong inconsistency between our toy fits and the reported DAMA results. We find that whilst the sign of $\hat{m}_j$ in the lower energy bins is still opposite to that reported in the DAMA fits, the amplitude is much smaller compared to our toy fits without the linear component, and we see much less difference in our fit results for the different $^3$H activity assumptions. This can be attributed to the linear component in each energy bin allowing for the deviation between observed and expected counts to be minimised with a much smaller fitted $\hat{m}_j$, given that $\hat{l}_j$ has the freedom to fit out differently for the different $^3$H activity cases. We conclude that this fitting methodology, in the presence of time-varying $^{210}$Pb and $^3$H backgrounds, cannot produce results consistent with those reported by DAMA, implying a feature in their data not captured by our toy model.

\begin{figure}[H]
    \includegraphics[scale=0.52]{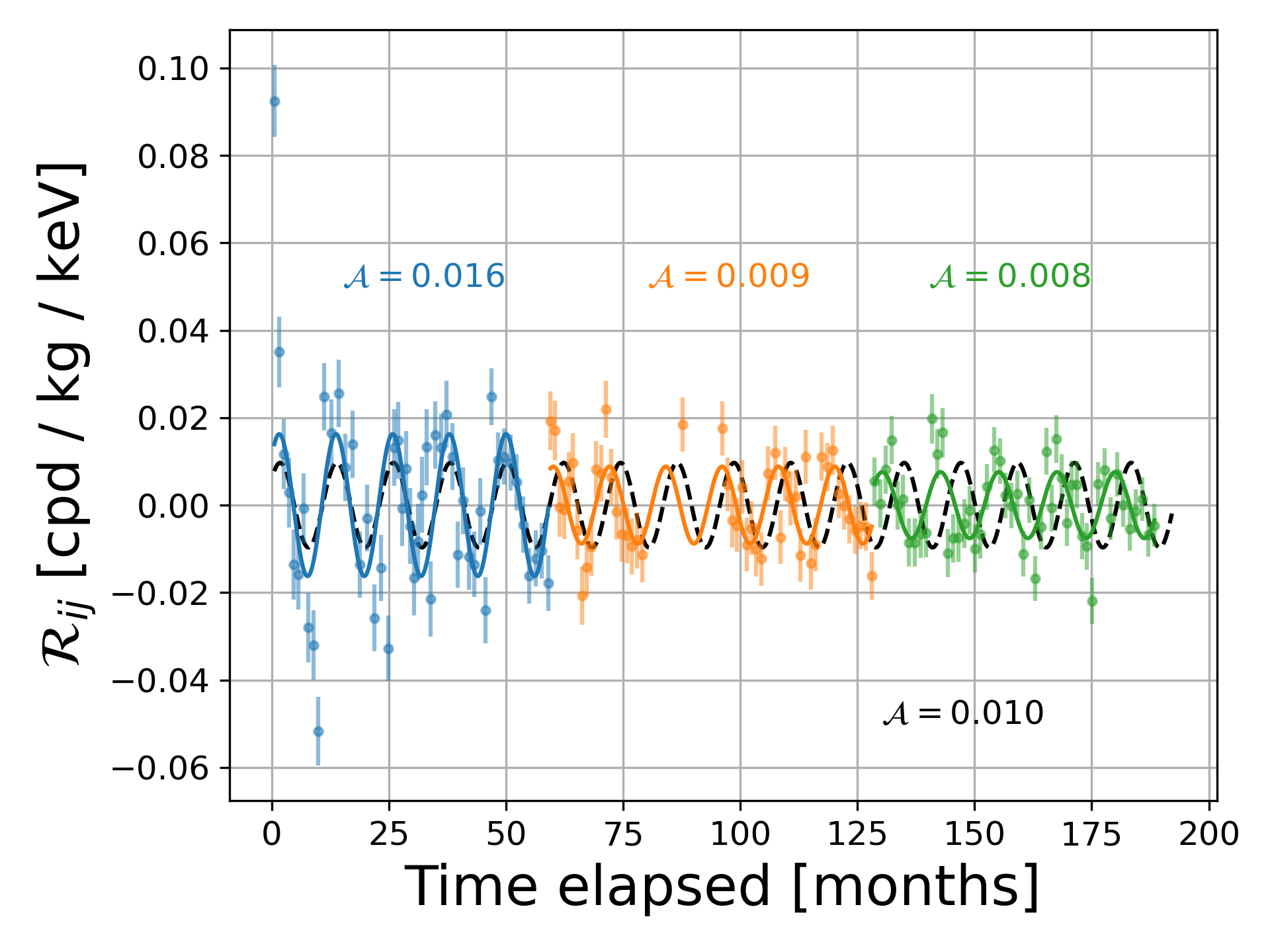}
    \caption{\label{fig:time_persistance}The result of performing the same fit as in Fig. \ref{fig:toy_dataset_residuals} to a single realisation of a toy DAMA/LIBRA dataset, this time partitioning the data into 3 equal time periods and performing the fit separately for each. As can be seen, the amplitude of the induced modulation decays over time. We compare with a fit to the dataset without any partitioning (dashed black).}
\end{figure}

An especially curious feature of the DAMA/LIBRA modulation signal is its persistence over time. Such behaviour is cited by DAMA when they argue that the modulation of their residual rate -- seen to be constant across the entirety of phases 1 and 2 of the experiment \cite{Bernabei_2021} -- could be due to the scattering of galactic dark matter. Clearly, though, the induced modulation effect coming from the presence of an exponentially falling background rate produces a modulation-like signature that is not constant in time. This is made evident in Fig. \ref{fig:time_persistance}, where the same fitting procedure is performed as in Fig. \ref{fig:toy_dataset_residuals}, this time fitting within 3 slices of time across the full exposure. It is clear that the amplitude of the fit is lower for later slices. It seems difficult, therefore, to attribute the DAMA signal to an induced modulation coming from exponentially decaying background activity.

\begin{figure}[H]
    \includegraphics[scale=0.55]{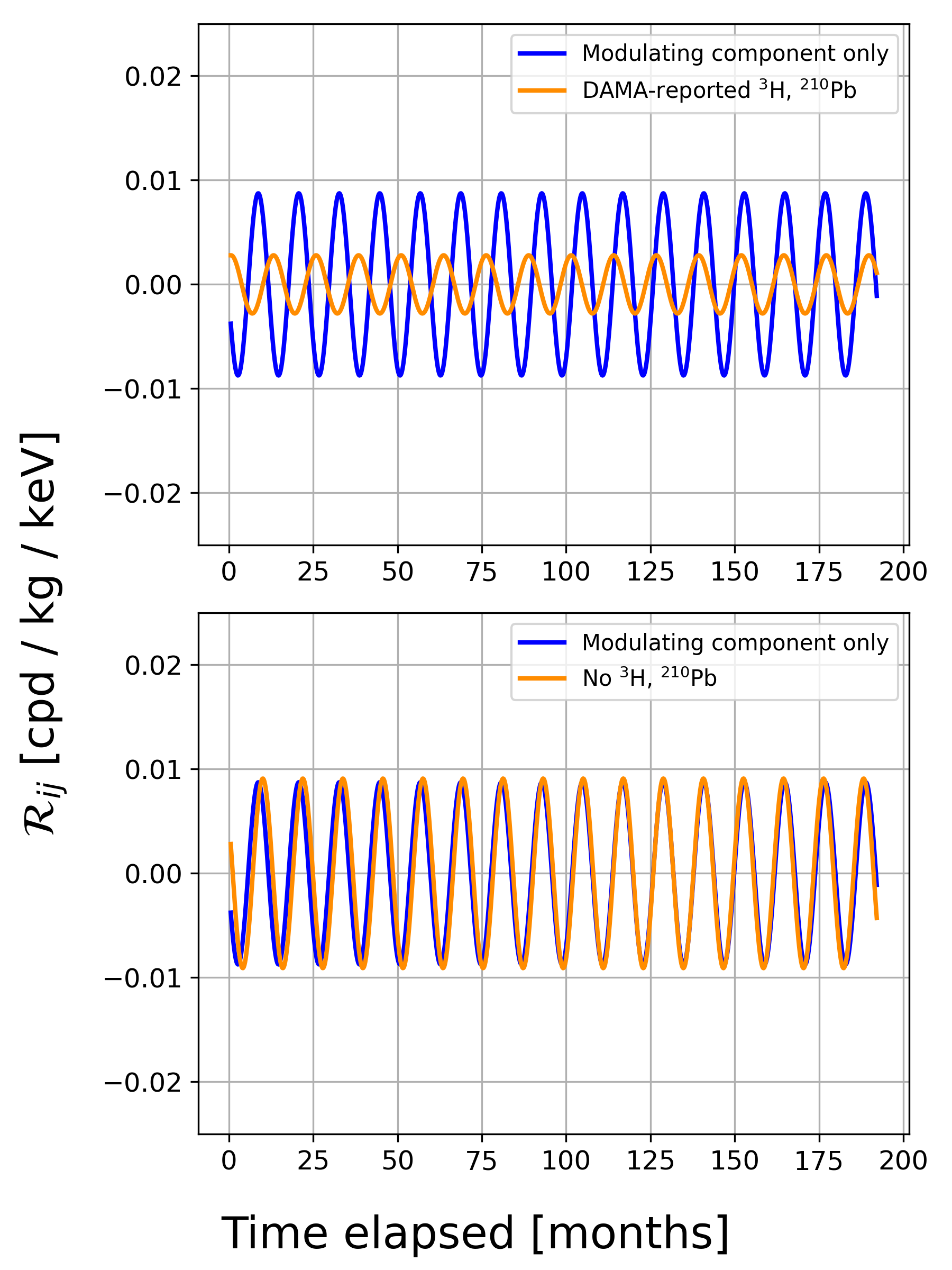}
    \caption{\label{fig:injected_signal}The result of performing the same fit as in Fig. \ref{fig:toy_dataset_residuals} to a single realisation of a toy DAMA/LIBRA dataset, but this time including a component with a genuinely oscillating event rate, as described in the text. We compare a fit to only the modulating component of the datset (blue), versus the full dataset (orange), under two assumptions on the initial activities of $^3$H and $^{210}$Pb. \textbf{Top:} initial activities of $^3$H and $^{210}$Pb are set to those reported by DAMA. \textbf{Bottom:} initial activities of $^3$H and $^{210}$Pb are set to zero.}
\end{figure}

To go one step further, we can conclude that not only is this induced modulation effect likely not the source of the signal, but that such an effect,  if present at a visible level in the DAMA data, would mask a truly modulating component. To make this clear, we extend our toy model to include a component with a truly modulating event rate. For simplicity, we give this component a flat energy spectrum within the ROI, predicting an expected 10$^6$ counts in the full DAMA/LIBRA exposure, and with a modulation amplitude that, in the absence of any other time-varying background components, would give a residuals distribution consistent with the DAMA result. In Fig.~\ref{fig:injected_signal} we show the result of the residuals fit in the case of the DAMA-reported upper limits on the activity of $^3$H and $^{210}$Pb, as well as a case where we set the activity of these two components -- which contribute most to the fall in activity over the lifetime of the experiment -- to zero. For comparison, we show in each case the result of a residuals fit to the genuinely modulating events only.

We find that the induced modulation effect acts against the true underlying modulating signal. When the activities of $^3$H and $^{210}$Pb are high enough to give a strong induced modulation effect, the best-fit modulation parameters are such that early on, the fitted signal is out of phase with the truly modulating component and with a lower amplitude. At later times, where the induced modulation effect is weaker, the fitted signal is in phase with the truly modulating component, again with a lower amplitude. This reflects that the induced modulation -- which is approximately $\pi$ out of phase with the best-fit result reported by DAMA -- is strong enough to fully nullify the true modulation at the initial times when it is strongest, and to produce a non-zero $\pi$-shifted modulation on top of this. At later times, when the induced modulation amplitude is much lower, we see that it acts to reduce the amplitude of the truly modulating component, but not to the same degree. In the case where the activities of $^3$H and $^{210}$Pb are zero, where the induced modulation effect is much weaker, we see that the true underlying modulation is almost entirely preserved. The constancy of the DAMA modulation suggests that the induced modulation effect must be negligible in the DAMA dataset, which leads us to conclude that the upper limits on $^3$H and $^{210}$Pb activity reported by the collaboration are almost certainly overestimates. If these activities were as high as the upper limit reported in Ref.~\cite{Bernabei_2008_2}, the induced modulation effect would act to give a time variation to the oscillations in the DAMA residuals distribution.

In summary, we have examined in detail the hypothesis that the results reported by DAMA/LIBRA could have arisen from a bias introduced by their particular analysis approach. We have found that this `induced modulation' is almost in antiphase with the signal reported by DAMA, given their choice in periods over which to take an average for their background subtraction. We use publicly available information to calculate our own upper limit estimation of the initial activity of the radiogenic background isotope $^3$H in the DAMA crystals, which is substantially lower than the upper limit reported by DAMA. Repeating our analysis with this alternative background model, we find that the amplitude of the induced modulation is also in disagreement with that reported by DAMA, by a factor $\sim$2. We have also examined the impact of a time-varying background on an alternative fitting methodology adopted by DAMA, once again showing through inconsistency with the DAMA result that this induced modulation effect cannot explain the DAMA signal. Finally, we have shown explicitly that the induced modulation coming from an exponentially decaying background presents as an oscillation in the residuals distribution with decaying amplitude, in stark contrast to the consistency of the oscillation characteristics reported by DAMA. We have shown that such an induced modulation would affect the way that a genuinely modulating component would present in the residuals distribution,  when such a genuine modulation is consistent with that reported by DAMA. We therefore conclude that the activities of radioisotopes such as $^3$H and $^{210}$Pb must be lower than the upper limits reported by DAMA, to the point where the induced modulation effect does not affect the consistent modulation seen in their residuals distribution. This conclusion holds regardless of the true origin of the underlying modulation; be it from galactic dark matter, or otherwise. Other work has already explored the possibility that an induced modulation of constant amplitude and with a phase consistent with that reported by DAMA could arise from a constantly and linearly increasing background rate over the runtime of the entire experiment \cite{Messina_2020}. However, we believe that the presence of such a contribution is unlikely, as it seems unphysical. We find that the mystery of the DAMA/LIBRA signal remains, and it is up to current experiments to provide answers. In particular, the SABRE project offers a unique opportunity to do this using crystals with a lower background rate than those used in the DAMA/LIBRA experiment, as well as a uniquely placed Southern Hemisphere detector.

\begin{acknowledgments}
The SABRE South program is supported by the Australian Government through the Australian Research Council (Grants: CE200100008, LE190100196, LE170100162, LE160100080, DP190103123, DP170101675, LP150100705). This research was partially supported by Australian Government Research Training Program Scholarships and Melbourne Research Scholarships. This research was supported by The University of Melbourne’s Research Computing Services and the Petascale Campus Initiative. We thank the SABRE North collaboration for their contribution to the SABRE South experiment design and to the simulation framework. We also thank the Australian Nuclear Science and Technology Organisation for the assistance with the material screening and the measurement of background radiation at SUPL.
\end{acknowledgments}

\nocite{*}

\bibliography{apssamp}

\end{document}

%% file: author_list.tex
\author{R.~S.~James}
\email{robert.james1@unimelb.edu.au}
\affiliation{School of Physics, The University of Melbourne, Victoria 3010, Australia}
\affiliation{ARC Centre of Excellence for Dark Matter Particle Physics, Australia}

\author{K.~J.~Rule}
\affiliation{School of Physics, The University of Melbourne, Victoria 3010, Australia}
\affiliation{ARC Centre of Excellence for Dark Matter Particle Physics, Australia}

\author{E.~Barberio}
\affiliation{School of Physics, The University of Melbourne, Victoria 3010, Australia}
\affiliation{ARC Centre of Excellence for Dark Matter Particle Physics, Australia}

\author{V.~U.~Bashu}
\affiliation{Department of Nuclear Physics and Accelerator Applications, The Australian National University, Canberra, ACT 2601, Australia}
\affiliation{ARC Centre of Excellence for Dark Matter Particle Physics, Australia}

\author{L.~J.~Bignell}
\affiliation{Department of Nuclear Physics and Accelerator Applications, The Australian National University, Canberra, ACT 2601, Australia}
\affiliation{ARC Centre of Excellence for Dark Matter Particle Physics, Australia}

\author{I.~Bolognino}
\affiliation{Department of Physics, The University of Adelaide, Adelaide, SA 5005, Australia}
\affiliation{ARC Centre of Excellence for Dark Matter Particle Physics, Australia}

\author{G.~Brooks}
\affiliation{School of Engineering, Swinburne University of Technology, Hawthorn, VIC 3122, Australia}
\affiliation{ARC Centre of Excellence for Dark Matter Particle Physics, Australia}

\author{S.S.~Chhun}
\affiliation{School of Physics, The University of Melbourne, Victoria 3010, Australia}
\affiliation{ARC Centre of Excellence for Dark Matter Particle Physics, Australia}

\author{F.~Dastgiri}
\affiliation{Department of Nuclear Physics and Accelerator Applications, The Australian National University, Canberra, ACT 2601, Australia}
\affiliation{ARC Centre of Excellence for Dark Matter Particle Physics, Australia}

\author{A.~R.~Duffy}
\affiliation{Centre for Astrophysics and Supercomputing, Swinburne University of Technology, Hawthorn, VIC 3122, Australia}
\affiliation{ARC Centre of Excellence for Dark Matter Particle Physics, Australia}

\author{M.~Froehlich}
\affiliation{Department of Nuclear Physics and Accelerator Applications, The Australian National University, Canberra, ACT 2601, Australia}
\affiliation{ARC Centre of Excellence for Dark Matter Particle Physics, Australia}

\author{T.~M.~A.~Fruth}
\affiliation{School of Physics, The University of Sydney, NSW 2006 Camperdown, Sydney, Australia}
\affiliation{ARC Centre of Excellence for Dark Matter Particle Physics, Australia}

\author{G.~Fu}
\affiliation{School of Physics, The University of Melbourne, Victoria 3010, Australia}
\affiliation{ARC Centre of Excellence for Dark Matter Particle Physics, Australia}

\author{G.~C.~Hill}
\affiliation{Department of Physics, The University of Adelaide, Adelaide, SA 5005, Australia}
\affiliation{ARC Centre of Excellence for Dark Matter Particle Physics, Australia}

\author{K.~Janssens}
\affiliation{Department of Physics, The University of Adelaide, Adelaide, SA 5005, Australia}
\affiliation{ARC Centre of Excellence for Dark Matter Particle Physics, Australia}

\author{S.~Kapoor}
\affiliation{School of Physics, The University of Sydney, NSW 2006 Camperdown, Sydney, Australia}
\affiliation{ARC Centre of Excellence for Dark Matter Particle Physics, Australia}

\author{G.~J.~Lane}
\affiliation{Department of Nuclear Physics and Accelerator Applications, The Australian National University, Canberra, ACT 2601, Australia}
\affiliation{ARC Centre of Excellence for Dark Matter Particle Physics, Australia}

\author{K.~T.~Leaver}
\affiliation{Department of Physics, The University of Adelaide, Adelaide, SA 5005, Australia}
\affiliation{ARC Centre of Excellence for Dark Matter Particle Physics, Australia}

\author{P.~McGee}
\affiliation{Department of Physics, The University of Adelaide, Adelaide, SA 5005, Australia}
\affiliation{ARC Centre of Excellence for Dark Matter Particle Physics, Australia}

\author{L.~J.~McKie}
\affiliation{Department of Nuclear Physics and Accelerator Applications, The Australian National University, Canberra, ACT 2601, Australia}
\affiliation{ARC Centre of Excellence for Dark Matter Particle Physics, Australia}

\author{P.~C.~McNamara}
\affiliation{Department of Nuclear Physics and Accelerator Applications, The Australian National University, Canberra, ACT 2601, Australia}
\affiliation{ARC Centre of Excellence for Dark Matter Particle Physics, Australia}

\author{J.~McKenzie}
\affiliation{School of Physics, The University of Melbourne, Victoria 3010, Australia}
\affiliation{ARC Centre of Excellence for Dark Matter Particle Physics, Australia}

\author{W.~J.~D.~Melbourne}
\affiliation{School of Physics, The University of Melbourne, Victoria 3010, Australia}
\affiliation{ARC Centre of Excellence for Dark Matter Particle Physics, Australia}

\author{M.~Mews}
\affiliation{School of Physics, The University of Melbourne, Victoria 3010, Australia}
\affiliation{ARC Centre of Excellence for Dark Matter Particle Physics, Australia}

\author{L.~J.~Milligan}
\affiliation{School of Physics, The University of Melbourne, Victoria 3010, Australia}
\affiliation{ARC Centre of Excellence for Dark Matter Particle Physics, Australia}

\author{J.~Mould}
\affiliation{Centre for Astrophysics and Supercomputing, Swinburne University of Technology, Hawthorn, VIC 3122, Australia}
\affiliation{ARC Centre of Excellence for Dark Matter Particle Physics, Australia}

\author{F.~Nuti}
\affiliation{School of Physics, The University of Melbourne, Victoria 3010, Australia}
\affiliation{ARC Centre of Excellence for Dark Matter Particle Physics, Australia}

\author{F.~Scutti}
\affiliation{Centre for Astrophysics and Supercomputing, Swinburne University of Technology, Hawthorn, VIC 3122, Australia}
\affiliation{ARC Centre of Excellence for Dark Matter Particle Physics, Australia}

\author{Z.~Slavkovsk\'{a}}
\affiliation{Department of Nuclear Physics and Accelerator Applications, The Australian National University, Canberra, ACT 2601, Australia}
\affiliation{ARC Centre of Excellence for Dark Matter Particle Physics, Australia}

\author{N.~J.~Spinks}
\affiliation{Department of Nuclear Physics and Accelerator Applications, The Australian National University, Canberra, ACT 2601, Australia}
\affiliation{ARC Centre of Excellence for Dark Matter Particle Physics, Australia}

\author{O.~Stanley}
\affiliation{School of Physics, The University of Melbourne, Victoria 3010, Australia}
\affiliation{ARC Centre of Excellence for Dark Matter Particle Physics, Australia}

\author{A.~E.~Stuchbery}
\affiliation{Department of Nuclear Physics and Accelerator Applications, The Australian National University, Canberra, ACT 2601, Australia}
\affiliation{ARC Centre of Excellence for Dark Matter Particle Physics, Australia}

\author{B.~Suerfu}
\affiliation{International Center for Quantum-field Measurement Systems for Studies of the Universe and Particles (QUP), High Energy Accelerator Research Organization (KEK), Oho 1-1, Tsukuba, Ibaraki 305-0801, Japan}

\author{G.~N.~Taylor}
\affiliation{School of Physics, The University of Melbourne, Victoria 3010, Australia}
\affiliation{ARC Centre of Excellence for Dark Matter Particle Physics, Australia}

\author{P.~Urquijo}
\affiliation{School of Physics, The University of Melbourne, Victoria 3010, Australia}
\affiliation{ARC Centre of Excellence for Dark Matter Particle Physics, Australia}

\author{A.~G.~Williams}
\affiliation{Department of Physics, The University of Adelaide, Adelaide, SA 5005, Australia}
\affiliation{ARC Centre of Excellence for Dark Matter Particle Physics, Australia}

\author{Y.~Xing}
\affiliation{School of Physics, The University of Melbourne, Victoria 3010, Australia}
\affiliation{ARC Centre of Excellence for Dark Matter Particle Physics, Australia}

\author{Y.~Y.~Zhong}
\affiliation{Department of Nuclear Physics and Accelerator Applications, The Australian National University, Canberra, ACT 2601, Australia}
\affiliation{ARC Centre of Excellence for Dark Matter Particle Physics, Australia}

\author{M.~J.~Zurowski}
\affiliation{School of Physics, The University of Melbourne, Victoria 3010, Australia}
\affiliation{ARC Centre of Excellence for Dark Matter Particle Physics, Australia}
\affiliation{Department of Physics, The University of Toronto, ON M5R 2M8, Canada}

\collaboration{The SABRE South Collaboration}